\documentclass[aps,pre,floats,superscriptaddress,twocolumn]{revtex4}
\usepackage{amssymb,amsmath}
\usepackage{amsmath,amssymb}
\usepackage{graphicx}
\usepackage{psfrag,textcomp}
\usepackage{color}
\usepackage[normalem]{ulem}

\def\beq{\begin{equation}}
\def\eeq{\end{equation}}
\def\bea{\begin{eqnarray}}
\def\eea{\end{eqnarray}}
\begin{document}
\title{Phases and fluctuations   in a model for asymmetric inhomogeneous fluid membranes}
\author{Niladri Sarkar}\email{niladri.sarkar@saha.ac.in}
\author{Abhik Basu}\email{abhik.basu@saha.ac.in}
\affiliation{Condensed Matter Physics Division, Saha Institute of
Nuclear Physics, Calcutta 700064, India}
\date{\today}

\begin{abstract}
We propose and analyze a model for phase transitions in an
inhomogeneous fluid membrane, that couples local composition with
curvature nonlinearly. For asymmetric membranes, our model shows
generic non-Ising behavior and the ensuing phase diagram displays
either a first- or a second-order phase transition through a
critical point (CP) or a tricritical point (TP), depending upon the
bending modulus. It predicts generic  nontrivial enhancement in
fluctuations of asymmetric membranes that scales with system size in
a power law fashion at the CP and TP in two dimensions,  not
observed in symmetric membranes. It also yields two-dimensional
Ising universality class for symmetric membranes, in agreement with
experimental results.

\end{abstract}
 \maketitle

\section{Introduction}
 Recent experiments on  giant
plasma membrane vesicles (GPMVs) isolated from living cells and
model membranes (e.g., artificially prepared lipid bilayers made of
lipids and cholesterol) display lateral inhomogeneities over length
scales much larger than the typical sizes of the constituent lipid
and protein molecules and stay in coexisting liquid-ordered ($L_o$)
and liquid-disordred ($L_d$) phases \cite{hetero,baum,vet1}. The
universal scaling exponents that characterize the miscibility
transition in GPMVs or model lipid bilayers are experimentally found
to be close to those of the two-dimensional (2D) Ising model
\cite{crit,honerkamp}. Ref.~\cite{ehrig}, by using lattice-based
Monte-Carlo simulations, studies phase separations and critical
fluctuations in two-component lipid membranes, and discusses the
possibility of both quasi-abrupt transitions (first order
transitions) and  second order transitions, depending upon the
details of lipid compositions. Plasma membranes of living cells are
however asymmetric \cite{alberts}. This asymmetry may affect the
macroscopic properties at the critical point \cite{asym}, although
the knowledge about it is still at a preliminary level \cite{sym1}.
Hence, in the absence of detailed experimental results on the
possible universal scaling properties of phase transitions in
asymmetric membranes, studies of phase transitions in asymmetric
membranes theoretically by using models with definite symmetries
should be welcome. In particular, studies on how a generic
curvature-composition nonlinearity, allowed by specific symmetry
considerations, can affect the resulting phase  behavior of the
model, should be of interest.

In this article we ask whether asymmetry in an inhomogeneous fluid
membrane can significantly enhance fluctuations near phase
transitions. Since any purported effects of asymmetry are likely to
depend on how asymmetry affects the system free energy, we
theoretically propose and study a model belonging to a particular
symmetry for phase transitions in both symmetric and asymmetric
inhomogeneous membranes. Our model is a reduced model, in which
the bilayer nature is not kept explicitly; instead it is replaced
effectively by an inhomogeneous single layer for simplicity. Our
main results are that depending upon the bending modulus $\kappa$
the composition field that describes the inhomogeneity of the
asymmetric membrane in our model may display (i) both first and
second order phase transition through a critical point (CP) and (ii)
a tricritical point (TP). In addition, our model predicts {\em
generic nontrivial enhancement} of fluctuations of asymmetric
inhomogeneous membranes near TP and CP with a variance that depends
on the system size $L$ in a power law fashion in two dimensions.
Further in our model, it is necessarily of second order belonging to
the 2D Ising universality class for symmetric membranes. We provide
a mean-field (MF) analysis of the phase diagram of the model and
calculate the universal scaling exponents that characterize the CP
and TP in our model by using perturbative renormalization group (RG) methods. Unlike more commonly used linear
curvature-composition coupling \cite{linear}, our model involves
non-linear curvature-composition coupling belonging to a particular
symmetry (see below). Thus it should be useful in understanding the
general implications of non-linear curvature-composition
relationships in experiments on membranes \cite{baum-nonlin}.
Validity of our model and the results that follow may be tested by
measuring fluctuations in asymmetric inhomogeneous membranes and
used to contrast with symmetric membranes. From perspectives of
theoretical physics, our model is an example of a fluid membrane
with internal degrees of freedom that displays rich critical
behavior.
The rest of the article is organized as follows: In Sec.~\ref{model}
we construct our model. We then discuss the mean-field analysis of
our model in Sec.~\ref{mft}, following by enumeration of the scaling
exponents in Sec.~\ref{scale}. We conclude and summarize our results
in Sec.~\ref{conclu}.

\section{Construction of the model}\label{model}

A real asymmetric lipid bilayer generally consists of two lipid
monolayers, which are constitutionally different (i.e.,
non-identical), e.g., cell membranes of eukaryotic
cells~\cite{alberts}. There are microscopic processes (e.g.,
flip-flop~\cite{alberts}) which lead to exchange of the constituent
molecules between the two monolayers of a bilayer. However, these
exchange processes are typically very slow. In addition, there are
biochemical processes which may favor maintaining compositional
differences across the leaflets~\cite{enzyme}. Thus, the
compositions of the two leaflets over relevant physical and
biological time scales may be very different. Coarse-grained models
for an asymmetric lipid bilayer thus require two sets of scalar
fields, one each for each of the monolayers, representing the
compositional degree of freedom in the corresponding monolayer, see,
e.g., Refs.~\cite{bi} for details. We, however, use a simpler
approach, where a single field $\phi$ is used together with a height
field to mathematically describe an asymmetric membrane. Here, the
asymmetry is no longer described by the composition differences
between the two monolayers, but rather by the breakdown of the
inversion symmetry of the membrane. Thus, effectively, only a single
layer is considered instead of a bilayer structure. There is already
a substantial body of work that uses a single composition field
degree of freedom, together with a height field, to model an
asymmetric inhomogeneous membrane; see, e.g., Refs.~\cite{linear}.
Our model is in the same spirit as them.

Here are the details of our model: We use a coarse-grained description to build our
model. For simplicity, we ignore the bilayer structure, and model it
 implicitly by a single inhomogeneous fluid membrane,
microscopically made up of two components (say, two types of
lipids), $A$ and $B$, with local concentrations $n_A$ and $n_B$,
 with vanishing surface tension
\cite{surf}, described by a height field $h(x,y)$ in the Monge gauge
\cite{weinberg-book} and the  concentration difference field $\phi
\equiv [n_A ({\bf x}) -n_B({\bf x})]$, as the relevant compositional
degree of freedom, since we are interested in phase transitions
in the membrane \cite{leib1}, and ${\bf x}=(x,y)$. The detailed
form of the free energy functional depends sensitively on the
underlying symmetry of the model system. We choose $\phi$ to have
the {\em Ising-symmetry},
interacting via local ferromagnetic Ising-like interactions~\cite{vet1}. 
Further, the system being asymmetric is {\em not} invariant
under the inversion symmetry $h\rightarrow -h$. Our imposition of
these symmetries for an asymmetric membrane, dictates that the
system and hence the corresponding free energy functional $\mathcal
F$ should be invariant under $(\phi,h)\rightarrow (-\phi,h)$
(hereafter SYMI). Thus in our model, one may have symmetry-allowed
{\em curvature-composition coupling} of the form $\lambda\phi^2
\nabla^2 h +\lambda_1 \phi^2 (\nabla^2 h)^2$. In contrast, symmetric
membrane models are typically invariant separately under
$\phi\rightarrow -\phi$ and $h\rightarrow -h$ (SYMII). Therefore,
taking everything into consideration $\mathcal F$ is
 \bea
  &&{\mathcal{F}}=\int d^dx [{r \over 2}\phi^2 + {b
\over 2}(\nabla \phi)^2 + {u \over 4!}\phi^4 + {v \over
6!}\phi^6\nonumber \nonumber \\ &+&{\kappa \over 2}(\nabla^2h)^2 +
\lambda \phi^2\nabla^2h + \lambda_1 \phi^2 (\nabla^2h)^2],
\label{free}
 \eea
 where $r\sim(T-T_c)$ with $T_c$ as the MF critical temperature for
$\phi$, and taking $\phi$ (having the dimension of concentration) to scale as $1/\xi_0^2$, 
$b\sim K_BT \xi_0^4$ (in 2D) is a measure of
energy-scale associated with composition fluctuations, where $K_B$
is the Boltzmann constant and $\xi_0$ is (microscopic) correlation
length of the order of a few angstroms. Coupling constants $u>0$
and $v>0$ determine the strengths of lipid-lipid interactions;
$v\phi^6$-term is added for thermodynamic stability (see below). The
$\lambda$-term is related to a
 local {\em spontaneous curvature} $c_0(\phi)=-\lambda\phi^2/\kappa$
\cite{helfrich,wall}.   Alternatively, it may be viewed as a {\em
local fluctuation} in $T_c$, such that local critical temperature
$T_{cL}=T_c-2\lambda \nabla^2h$. Coupling constant $\lambda$
embodies asymmetry in the model and implicitly contains information
about geometry (e.g., shape, packing) and asymmetric distribution of
lipid molecules in the system. On physical grounds, following
the arguments in Ref.~\cite{leib1}, we set $\lambda n_0^2 \sim
\kappa H_0$, where $n_0\sim 1/\xi_0^2$ is a mean concentration and
$H_0\sim 10^{6}m^{-1}$ is a typical (molecular) spontaneous
curvature. This yields $\lambda\sim \kappa H_0\xi_0^4$. Since
$\xi_0$, the microscopic correlation length, should scale with the
linear size of the constituent lipid molecules, $\lambda$ may be
experimentally varied by considering samples made of lipids of
different molecular sizes. Of course, for a pure fluid membrane
$\phi=0$~\cite{udo-rev}. Notice that our model is not the most
generic in that a SYMI violating asymmetric linear
curvature-composition coupling term may be present. Effects of such
a coupling, i.e., $c_0(\phi)$ being proportional to $\phi$ with the
system invariant under $(\phi,h)\rightarrow (-\phi,-h)$ (hereafter
SYMIII), have been extensively studied \cite{linear}. Our choice of
curvature-composition coupling is made for the purpose of
illustration~\cite{foot7} and complementary to that in
Refs.~\cite{linear}.
Furthermore, the $\lambda_1$-term may be interpreted as a
contribution to $T_{cL}$ or to a local (effective) bending modulus
$\kappa_L= \kappa+2\lambda_1 \phi^2$. Here, we have omitted
the geometric nonlinearities, which are subdominant to the existing
nonlinear terms (the $u$- and $\lambda$-terms)
~\cite{weinberg-book,geom,geom3}. Note that the absolute sign of
$\lambda$ is arbitrary and may be switched by $h\rightarrow -h$. The
$\lambda$-term violates the inversion symmetry $h\rightarrow -h$;
where as for a symmetric membrane one has $\lambda=0$, leading to
invariance under SYMII.  Lastly, since our model effectively assumes
the bilayer to be a single incompressible sheet, it is unable to
capture the effects of coupling of the local bending and densities
of the two monolayers \cite{seifert1}, which yields a new slow mode
related to fluctuations in the density difference in the two
monolayers. Notice that in constructing our model we have been
guided solely by general symmetry considerations, without any
reference to microscopic or molecular level details. We believe it
is useful to theoretically construct models with specific symmetry
using this approach and study the ensuing macroscopic consequences,
in view of the relative lack of quantitative results on the
universal scaling properties of phase transitions in asymmetric
membranes, in comparison with their symmetric counterparts.

\section{Mean-field analysis}\label{mft}

\begin{figure}[htb]
\includegraphics[height=6cm]{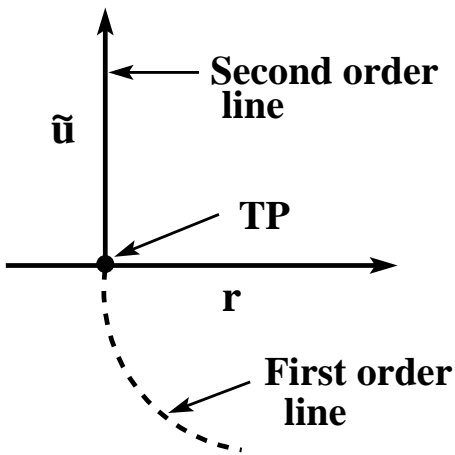}
\caption{Schematic mean-field phase diagram in the $r-\tilde u$
plane, depicting line of critical points, a TP and a first order
transition.} \label{first}
\end{figure}

 It is instructive to begin with a MF analysis for the phase
transitions in the model, ignoring the $\lambda_1$-term (it is
subdominant to the existing terms; see below). The MF is constructed
by ignoring all spatial correlations and in terms of constant values
of the order parameter field $\phi=m$ and the curvature $\nabla^2 h
=c$. Here, $c$ and $m$ may both be zero or non-zero depending upon
the phase concerned. The geometric nonlinearities are neglected.
First consider a symmetric membrane, for which within MF,
$c=\nabla^2 h=0$ for all $T$ and $\phi=m=0$ for $T\geq T_c$ and
$\phi=m\neq 0$ for $T<T_c$, respectively, thus yielding Ising-like
MF behavior with a second-order transition at $T_c$.  In contrast,
for an asymmetric membrane at the MF level for $T\geq T_c$ one
obtains, $c=\kappa\nabla^2 h = -\lambda\phi^2=-\lambda m^2=0$.
However, for $T<T_c$, one has $\phi = m \neq 0$, thus giving
$c=-\lambda m^2$, implying a non-zero radius of curvature for an
asymmetric membrane in the ordered phase. Ignoring the spatial
correlations of $\phi$ in the spirit of MF analyses, an effective
Landau free energy $\mathcal F_e$ for a constant $\phi=m$ may now be
constructed by replacing $c$. It has the form
 \begin{equation} {\mathcal
F_e}=\frac{r}{2}m^2 +\tilde u m^4 + vm^6,\label{effre}
\end{equation}
 where $\tilde u\equiv u-a\lambda^2/\kappa$ ($a>0$ is
a $O(1)$ numerical coefficient) can be $\tilde u$ positive, negative
or zero. Within the mean field picture, the effective Landau free
energy (\ref{effre}) may be analysed in a standard
way~\cite{chaikin}: one finds the following possibilities: (i) A CP
at $r=0$ with $\tilde u
>0$, (ii) a first order phase transition at $r=|\tilde u|^2/(2v)=|u-
a\lambda^2/\kappa|^2/(2v)$ for $\tilde u <0$, and (iii) a TP at
$r=0,\,\tilde u=0$, i.e., $u=a\lambda^2/\kappa$ as schematically
shown in Fig.~\ref{first}; this phase diagram is identical to that
observed in the normal-superfluid transition in liquid helium
mixtures~\cite{hel}. Evidently, from its definition as given above
$\tilde u
>0$ or $<0$ for large or small $\kappa$, respectively, implying
lipid domain formations proceed through second or first order
transitions for stiff (large $\kappa$) or soft (small $\kappa$)
membranes, respectively. We now critical exponents $\eta_\phi$ and
$\nu$ through the relations $\langle \phi({\bf x})\phi(0)\rangle
\sim |{\bf x}|^{2-d-\eta_\phi} f_\phi(|{\bf x}|/\xi_\phi)$ where $\xi_\phi\sim
|T-T_c|^{-\nu}$ is the correlation length. Thus, one finds from the
MF theory \cite{chaikin} $\eta_\phi=0$ and $\nu=1/2$ at both CP and
TP. The presence of the first order transition in this model clearly
underlines the requirement of a $v\phi^6\,(v>0)$ term in $\mathcal
F$ for thermodynamic stability.

We now complement our MF analysis above by analyzing the
properties of the fluid membrane in the spirit of MF. We start with
the partition function corresponding to the free energy functional
(\ref{free}):
\begin{equation}
{\mathcal Z}=\int {\mathcal D}\phi {\mathcal D}h \exp [-{\mathcal
F}/K_BT].\label{part}
\end{equation}
 By expanding perturbatively
and integrating $\phi$~\cite{int} (we ignore the $u$-term here), one
may further obtain for effective bending rigidity $\kappa_e = \kappa
- \lambda^2\int_{\bf q} \langle (\phi ({\bf q})\phi ({\bf
-q}))^2\rangle$ to the lowest order (setting $K_BT=1$), where $\bf
q$ is a wavevector. We shall see below that $\kappa_e$ becomes
scale-dependent, i.e., it depends on the local scale $l$. Since
$\langle (\phi ({\bf q})\phi ({\bf -q}))^2\rangle$ is positive
definite, $\kappa_e<\kappa$ generally, opening up the possibility of
$\kappa_e$ being zero or negative. Since $\kappa_e$ is generically
smaller than $\kappa$, the asymmetric membrane will be generically
softer or more flexible than a pure fluid membrane, a feature that
should also be displayed by models belonging to
SYMIII~\cite{linear}, which use
bilinear curvature-composition coupling. 
One may further define a persistence length $\xi_p$, such that
$\kappa_e(l=\xi_p)=0$. Noticing that for a system with linear
size $L$, the correction to $\kappa_e$ scales as $L^2$ in 2D at the
MF $T_c$ given by $r=0$, we find $\xi_p\sim [\kappa /(
\lambda^2)]^{1/2}$, thus $\xi_p\sim 1/\lambda$ for fixed $\kappa$
and $T$. Physically, this essentially gives a linear scale over
which the membrane remains flat. Clearly, in our model, for an
experimentally accessible asymmetric membrane with nearly flat
conformation, (bare) $\lambda$ should be sufficiently small, such
that $\xi_p\gg L$, linear dimension of the system. However, for
quantitative predictions on $\xi_p$, one requires to have a
numerical estimate of $\lambda$, which at present is lacking.
Construction of atomistic models and numerical studies on them
linking microscopic structures with macroscopic properties or
parameters (e.g., $\kappa,\,\lambda$) would be
useful~\cite{desserno},  which are beyond the scope of this work.
For $\kappa_e <0$, our model shows instabilities akin to
Refs.~\cite{linear}. Notice however that the instabilities here for
$\kappa_e <0$ are due to a nonlinear curvature-composition coupling,
unlike for models in Refs.~\cite{linear}, where bilinear
curvature-composition couplings are responsible for the
instabilities. In the remaining part of the article below, we only
consider $\kappa_e>0$. Furthermore, considering that $\kappa_e$
has got a $O(L^2)$ (perturbative) correction to its bare value
$\kappa$ coming from the curvature-composition interaction, thence
at 2D, with ${\bf n}={\boldsymbol \nabla} h$ as the local normal to
the membrane (to the leading order in smallness), the variance of
$\bf n$ has a (perturbatively obtained, ignoring logarithms) contribution
\begin{equation}
\Delta= \langle n^2 ({\bf x})\rangle \sim O(\lambda^2 L^2),
\label{vari1}
\end{equation}
in addition to the contribution from (bare) $\kappa$. Thus,
$\Delta$ formally diverges as $L\rightarrow\infty$. This is
significantly stronger than the well-known $\log L$ dependence of
$\Delta_s=\langle n^2 ({\bf x})\rangle$ that ensues for a pure
symmetric fluid membranes~\cite{weinberg-book}. Even for an
inhomogeneous symmetric membrane, $\Delta_s$ must have a weaker
$L$-dependence than $L^2$, since the dominant nonlinearity
responsible for the $L^2$-behavior (i.e., the $\lambda$-term) in our
model is always absent for a symmetric membrane. This is thus a
feature of an asymmetric membrane described by our model that may be
directly testable in experiments and can be contrasted with the same
from symmetric membranes.

\begin{figure}[htb]
\includegraphics[height=6cm]{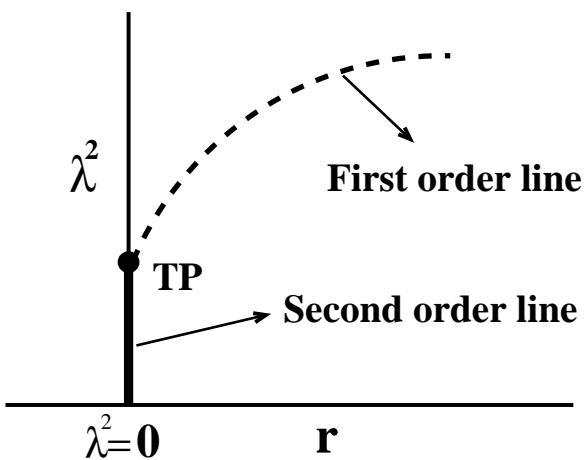}
\caption{Schematic mean-field phase diagram in the  $r-\lambda^2$
plane, depicting lines denoting first (broken line) and second
(thick continuous line) transitions and a TP.} \label{second}
\end{figure}

One may also present an equivalent phase diagram in the $r-\lambda$ plane: see
Fig.~\ref{second} for a schematic phase diagram in the $r-\lambda$ plane.

Compare now with linear curvature-composition models \cite{linear}
having a curvature-composition coupling of the form $\tilde \lambda
\phi\nabla^2 h$ (with $\lambda=0$). It can be shown in that (a)
$\tilde\lambda$ does not contribute to the fluctuation corrections
of $\phi$; with $u$ remaining as the most relevant nonlinearity,
large-scale properties of $\phi$ is identical to that of the  Ising
model, and (b) due to the absence of any nonlinear term there are no
non-trivial fluctuation corrections to $\kappa$ from
$\tilde\lambda$, and hence, as long as $k_e>0$, $h$ is identical to
that of a pure fluid membrane. In contrast, our model for asymmetric
membranes displays generic non-Ising behavior of $\phi$ including a
first order transition and scale-dependent diverging fluctuations of
$h$. Furthermore, as already pointed out above, for a symmetric
membrane with SYMII at 2D, $\langle n^2 ({\bf x})\rangle \sim \log
L$~\cite{weinberg-book}, much weaker than its $L^2$-dependence for
an asymmetric membrane in our model~\cite{geom1}.  We confirm below
the existence of CP and TP, extract the corresponding universal
scaling exponents and show the divergence of the variance of
asymmetric membrane fluctuations in a perturbative
RG calculation~\cite{whyrg}.  

\section{Scaling exponents near CP and TP}\label{scale}

For a symmetric membrane ($\lambda=0$) the remaining geometric
non-linearities are irrelevant at 2D in the presence of the $u$-term
with upper critical dimension $d_c=4$, and hence are ignored.
Effectively, thus in our model there are no {\em relevant} coupling
between $\phi$ and $h$ in a symmetric membrane. Consequently the
critical behavior of $\phi$ is identical to the Ising model, thus
belonging to the 2D Ising universality class, already studied
extensively in a RG framework \cite{amit,zinn}: One obtains
$\eta_\phi=\epsilon^2/54,\,\nu_\phi=1/(2-{\epsilon \over 3})$ with
$\epsilon=4-d>0$ (the $\phi^6$ interaction is irrelevant in a RG
sense). This is in agreement with experimental results on lipid
bilayers by several groups~\cite{vet1,honerkamp}. Further,
fluctuation statistics of $h$ is identical to that of a pure fluid
membrane. Assuming a nearly flat conformation (see
discussions in the concluding section), predictions from our model is more complicated
for an asymmetric membrane for which $\lambda\neq 0$ with $d_c=4$,
along with $u$ are relevant couplings (in a RG sense). Thus the
ensuing universal scaling behavior  at the physically relevant
dimension $d=2<d_c$ is expected to be different from both the
mean-field scaling behavior for an asymmetric membrane discussed
before and 2D Ising behavior of a symmetric membrane. In order to
 analyze this, we perform a systematic RG analysis of the model following standard procedure well-documented in the
literature
~\cite{amit,zinn}. 
 In particular, the renormalization $Z$-factors for the different vertex functions, which
are introduced to absorb ultra-violet (UV) divergences in the
theory, are calculated by using a dimensional regularization
together with a minimal subtraction scheme in terms of an arbitrary
momentum scale $\mu$~\cite{amit,zinn}. The vertex functions in
the present model, which have primitive divergences and hence
require renormalization, are given by (the right hand sides of the
expressions below are the bare values of the vertex functions in our
model that can be easily read off from the free energy functional
(\ref{free})).
\begin{eqnarray}
&&\frac{\delta^2
\Gamma}{\delta\phi_{\bf k}\delta\phi_{\bf -k}}=r+k^2,\\
&&\frac{\delta^2 \Gamma}{\delta h_{\bf k} \delta h_{\bf -k}}=\kappa
k^4,\\
&& \frac{\delta^3 \Gamma}{\delta \phi_{\bf q}\delta\phi_{\bf
-k+q}\delta h_{\bf
 k}}=
 2\lambda,\\
&&\frac{\delta^4\Gamma}{\delta \phi_{\bf k}\delta\phi_{\bf
q_1}\delta\phi_{\bf q_2}\delta\phi_{\bf -k-q_1-q_2}}=u,
\label{vertex}
\end{eqnarray}
where $\Gamma$ is the vertex generating functional and is the
Legendre transform of the free energy corresponding to the partition
function (\ref{part}); ${\bf k,q,q_1,q_2}$ are Fourier wavevectors.
The resulting RG flow equation yields non-trivial scaling of the
correlation functions and thermodynamic quantities at the RG fixed
points (FP). The relevant Feynman diagrams are shown in the Appendix
below.

\begin{figure}[htb]
\includegraphics[height=5.5cm]{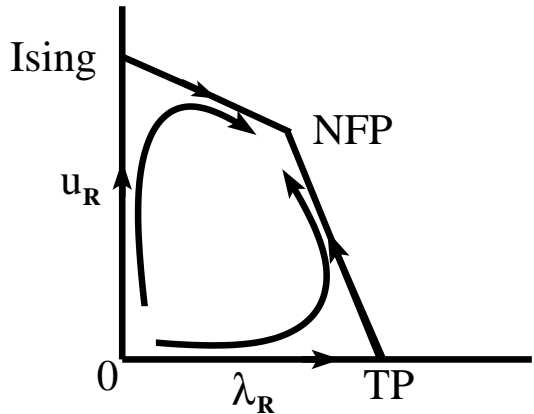}
\caption{A schematic stability diagram of FPs in the $\lambda_R-u_R$
plane. Arrows indicate stable directions (see text).}\label{flow}
\end{figure}

For calculational convenience we define an implicitly
scale-dependent bare coupling constant $g$ by ${g \over 4!} \equiv
{u \over 4!}- {8 \lambda^4\mu^{-\epsilon} \over 16\pi^2\epsilon}$,
such that the inhomogeneous part of the fluctuation corrections to
$u$ is absorbed and define renormalized coupling $u_R=Z_g{g \over
4!(4\pi)^{d/2}}\mu^{-\epsilon}$. We similarly define renormalized
$\lambda_R=Z_\lambda \lambda \mu^{-\epsilon/2}/(4\pi)^{d/4}$.
Defining the RG $\beta$-function by
$\beta_a=\mu\frac{\partial}{\partial\mu} a_R$, where
$a_R=u_R,\lambda_R$, we find 
 \bea
\beta_u &=& u_R[-\epsilon + 72u_R - 48\lambda_R^2],
\label{betau} \\
\beta_\lambda &=& \lambda_R[-{\epsilon \over 2} + 24u_R -
6\lambda_R^2 +  {192\lambda_R^4 \over \epsilon} ].
\label{betalambda} \eea
 The FPs are determined by the zeros of the
 $\beta$-functions above: (i) Gaussian (or trivial) fixed point given
 by $u_R=0=\lambda_R$, (ii) Ising FP given by
 $u_R=\epsilon/72,\,\lambda_R=0$, (iii) tricritical FP
 given by $u_R=0,\,\lambda_R^2=0.07\epsilon$ and (iv) nontrivial fixed
point (NFP) given by $u_R=0.02\epsilon,\,\lambda_R^2=0.013\epsilon$.
Very small value of $\lambda_R^2$ at NFP suggests larger
$\xi_p$ and hence easier experimental accessibility at NFP. Each
 of these FPs represents a different universality class, characterized by
 a different set of values of the critical exponents~\cite{loop}. For example,
 we find for
\begin{itemize}
 \item Gaussian FP: $\eta_\phi=0=\eta_h$, $1/\nu =
 2$,
 \item Ising FP: identical to the symmetric membrane case,
 \item TP: $\eta_\phi=0$, $\eta_h=-0.28\epsilon,
\,\, {1 \over \nu}= 2+0.179\epsilon$, and
\item NFP:
$\eta_\phi=0.04\epsilon^2,\, \eta_h=-0.05\epsilon,\, {1 \over \nu}=
2-0.3\epsilon$.
\end{itemize}
Thus, $\eta_h <0$ at all the FPs,
in agreement with our qualitative discussions above on
$\kappa_e<\kappa$. From the scaling exponents calculated above, we
find that
\begin{itemize}
\item
$C_\phi(|{\bf x}|)=\langle \phi ({\bf x})\phi (0)\rangle\sim |{\bf x}|^{2-d}$ at the
TP,
\item $C_\phi (|{\bf x}|)
\sim |{\bf x}|^{2-d-\epsilon^2/54}$ at the Ising FP,
\item $C_\phi (|{\bf x}|) \sim
|{\bf x}|^{2-d-0.04\epsilon^2}$ at the NFP, \end{itemize}
 where $\bf x$ is a
spatial separation vector in $d$-dimension. Thus,
$C_\phi(|{\bf x}|)$ varies most slowly with $|{\bf x}|$ at TP. Vanishing
$\eta_\phi$ at this order at TP is fortuitous and does not imply MF
result; $\nu$ picks up correction over its MF value already at this
order. Let us find out how the correlation of local normal ${\bf n}$
to the membrane surface behaves. Noting that $\langle h({\bf
x})h(0)\rangle \sim |{\bf x}|^{4-d-\eta_h}$, we find $\langle{\bf n}({\bf
x})\cdot {\bf n}(0)\rangle$ scales as $ \sim |{\bf x}|^{2-d+0.05\epsilon}$
at the NFP and $\sim |{\bf x}|^{2-d+0.28\epsilon}$ at the TP. Hence,
\begin{itemize}
\item At TP, $\Delta\sim L^{0.28\epsilon}$,
\item AT NFP,  $\Delta\sim L^{0.05\epsilon}$.
\end{itemize}
Compare this with a symmetric inhomogeneous membrane or a pure fluid
membrane. Disregarding the geometric nonlinearities (which are
irrelevant in a RG sense in the presence of $u$ and $\lambda$), for
a symmetric membrane variance $\Delta_s=\langle n^2({\bf x})\rangle$
 varies as $\log L$~\cite{geom1}. Thus, $\Delta$ depends on $L$ much
more strongly than $\Delta_s$. For example, measuring lengths in the unit of
molecular cut-off $\sim 10 nm$ and taking $L\sim 10\mu m$ (typical
size of an eukaryotic cell), we find $\Delta/\Delta_s\sim 10$ at TP
and 2 at NFP.
Finally, we consider linear stability of the different FPs.  We find
(i) the Gaussian FP $u_R=0,\lambda_R=0$ is unstable along both $u_R$
and $\lambda_R$ directions, (ii) the Ising FP  is unstable along the
$\lambda_R$ direction, but stable along the $u_R$-direction, (iii)
the TP is unstable along the $u_R$ direction but stable along the
$\lambda_R$ direction, and finally (iv) the NFP is stable along both
the directions. Not surprisingly, the TP can be reached only by
proper tuning of the parameters.  The RG flow lines are
schematically described in Fig.~\ref{flow}.

\section{Conclusion}\label{conclu}

In this article, we have proposed and studied a model, belonging to
a particular symmetry, for asymmetric inhomogeneous membranes. We
have analyzed the nature of phase transitions and fluctuations in
the model. Notice that in our model, the local spontaneous
curvature $c_0(\phi)=-\lambda\phi^2/\kappa$. Hence, the average
spontaneous curvature $\propto \langle \phi^2\rangle=m^2$ in MF, and
is zero in the high temperature disordered phase ($T>T_c$) and
non-zero in the low temperature ordered phase ($T<T_c$). However, if
we go beyond MF, then $\langle \phi^2\rangle$ is non-zero, due to
fluctuations, for both $T>T_c$ and $T<T_c$. Thus, when fluctuations
are taken into account, the model continues to have a non-zero
global spontaneous curvature $\overline C$, patrametrized by the
ratio $\lambda/\kappa$, at all $T$. The value of $\overline C$
generally depends on $T$, and is different in the disordered and
ordered phases, with a divergence (for a formally infinitely large
system) at $T_c$. Thus, our model provides a description for a
system with an overall curvature, which changes with $T$. For the 
validity of the calculational scheme employed here, one should have $\overline C L \ll 1$.
 Our model displays generic enhancement of fluctuations of
 asymmetric membranes
 near its second order phase transition, in contrast to symmetric membranes. For the latter, critical fluctuations
of $\phi$ are independent of $h$ and belongs to the 2D Ising
universality class, a feature observed in experiments on lipid
bilayers and GPMVs \cite{vet1,crit}. Our results for miscibility
transitions in asymmetric inhomogeneous membranes are much more
dramatic displaying generic non-2D Ising-like behavior, with  both
second and first order transitions and a TP. Further, asymmetric
membranes are found to have dramatically enhanced $L$-dependent
(formally diverging in the thermodynamic limit) fluctuations at TP
and CP in 2D, in contrast to symmetric membranes. Thus measurements
of membranes fluctuations near CP or TP are important, and should
yield signatures of asymmetry and composition-curvature coupling.
First order transitions in our model is noteworthy, since a similar
possibility has been discussed in Ref.~\cite{ehrig,baum-nonlin}.
Experiments on carefully prepared asymmetric membranes (see, e.g.,
Ref.~\cite{asym1})  by standard, e.g., fluorescence, methods (see,
e.g., Ref.~\cite{honerkamp}) should be useful, although more
complicated curvature-composition coupling \cite{baum-nonlin} may be
needed for quantitative reproduction of experimental data. At a
technical level, for a generic nonlinear coupling of the form
$f(\phi)\nabla^2 h$, where $f$ is an arbitrary function of $\phi$,
(in our case $f(\phi)\sim \phi^2$), there is a general possibility
of nontrivial renormalization of $\kappa$ via  nonlinear
contributions to the self-energy $\Sigma_{hh}$~\cite{amit} at
various orders of the underlying perturbation expansion. Our results
here is a specific example of such a possibility.  As mentioned
above, asymmetric inhomogeneous membrane models belonging to SYMIII
also display reduction in $\kappa_e$ due to composition fluctuations
and thus are qualitatively similar to our results.  In fact, far
away from $T_c$, models belonging to both SYMI and SYMIII yield
qualitatively similar behavior for membrane fluctuations (they
should, however, predict different phase diagrams for  $\phi$). But
their predictions for membrane fluctuations differ significantly as
$T\rightarrow T_c$ (i.e., near the critical zone), since
(considering $\kappa_e> 0$),  in the absence of any nonlinear term
that involves $h$ in models belonging to SYMIII, there are no
nontrivial (scale-dependent) fluctuation corrections to $\kappa$.
Hence, a linear curvature composition coupling model cannot lead to
any change in the scaling of the asymmetric membrane fluctuations
(although the magnitude can be enhanced). In contrast, due to the
non-trivial renormalization of $\kappa_e$, asymmetric membrane
fluctuations in our model displays  enhanced fluctuations near
$T_c$, with {\em non-trivial corrections} to the scale dependences
of the fluctuations. In a recent study on symmetric membranes by
Ayton {\em et al}~\cite{ayton}, the curvature composition coupling
is of the form $\phi (\nabla^2 h)^2$, which breaks the Ising
symmetry. It can be shown, by integrating the composition field,
that $\kappa_e < \kappa$ in the model of Ref.~\cite{ayton}, in the
model leading to enhancement of membrane fluctuations, qualitatively
similar to our results here. Thus, our study may be viewed as an
asymmetric membrane analog of Ref.~\cite{ayton}, where the Ising
symmetry is kept, but the inversion symmetry is broken.

Direct  comparisons of our results with experiments on inhomogeneous
asymmetric bilayer is difficult, mainly due to the reduced nature of
our model. Nonetheless, whether or not mechanisms as discussed in
our model is operative in a specific experimental set up, e.g.,  our
predictions on the connection of $\kappa$ with the order of
transitions and large fluctuations of asymmetric membranes near TP
or CP, may be tested by measuring membrane fluctuations.
Complementing our coarse-grained modeling by numerical studies on
microscopic models linking microscopic structures with macroscopic
properties (e.g., $\kappa,\lambda$) would be useful in the present
context~\cite{desserno}.  Since the signature and magnitude of the
effective coupling $\tilde u$ depends directly on
$\lambda^2/\kappa$, our system may be tuned by controlling the
membrane bending stiffness $\kappa$. Alternatively, as
discussed earlier, since $\lambda\sim H_0\xi_0^4\sim \xi_0^3$
(taking $H_0\sim \xi_0^{-1}$, although $H_0^{-1}>\xi_0$ generally),
$\lambda$ can be tuned by considering lipid membranes with lipid
molecules having different linear sizes, and hence with different
$\xi_0$. Considering the strong dependence of $\lambda$ on $\xi_0$,
this should be a promising route. Asymmetric membranes may be
prepared and their phase behavior investigated by combining the
Langmuir-Blodgett/Schaefer method~\cite{lang} with
fluorescence-based imaging. Membrane fluctuation measurements may be
done by spectroscopic methods \cite{flicker}. Bending modulus
$\kappa$ may be tuned experimentally, e.g., by cholesterol or BAR
proteins~\cite{exp}.
 Accessing TP
experimentally is expected to be considerably
more difficult due to the additional tuning required. 
Since $\kappa_e <\kappa$ generally, for a sufficiently large
membrane a first order transition may take place, depending upon the
value of $\kappa$. .
%
  Since {\em in-vivo} membranes (e.g., red blood cell membranes)
tend to have small but finite shear modulii, it will be interesting
to theoretically investigate effects of interactions between
in-plane displacements and compositional degrees of freedom. Since
the divergence of asymmetric membrane fluctuations occurs exactly at
CP or TP,  it will also be interesting to study the possibility of
budding transitions in our model and  if the diverging fluctuations
are signatures of the budding dynamics, as discussed in
Ref.~\cite{kumar}. Lastly, as we mentioned above, a limitation of
our model is the lack of bilayer structure in it. This, as
elucidated in Ref.~\cite{seifert1},  may affect the macroscopic
behavior. This can be incorporated by generalizing our single
compositional field description to two such fields. Both of these
fields should be mutually interacting, representing the interactions
between the monolayers, in addition to their couplings with $h$,
which should be of the form allowed in SYMI. Analyses of this
generalized model, although more complicated, can be done in a
straight forward way as here. Qualitative features found in our
model should be preserved, in addition to possible emergence of new
features, e.g., multicritical points and  additional slow
mode~\cite{seifert1} etc. We hope our results will stimulate further
theoretical and experimental works along these directions.

\section{Acknowledgement} AB thanks the Max-Planck-Gesellschaft (Germany) and the
Department of Science and Technology/Indo-German  Science and
Technology Centre (India) for partial financial support through the
Partner Group programme (2009).

\section{Appendix}
Here we show all the relevant one- and two-loop Feynmann diagrams
for our model. The internal continuous line denote $\langle\phi({\bf
q})\phi(-{\bf q})\rangle$ correlator. The internal broken line
denote $\langle h({\bf q}h(-{\bf q}))\rangle$ correlator. The
external continuous line stands for $\phi$ and the external broken
line stands for $h$.
\begin{widetext}

\begin{figure}[htb]
\includegraphics[height=7.0cm]{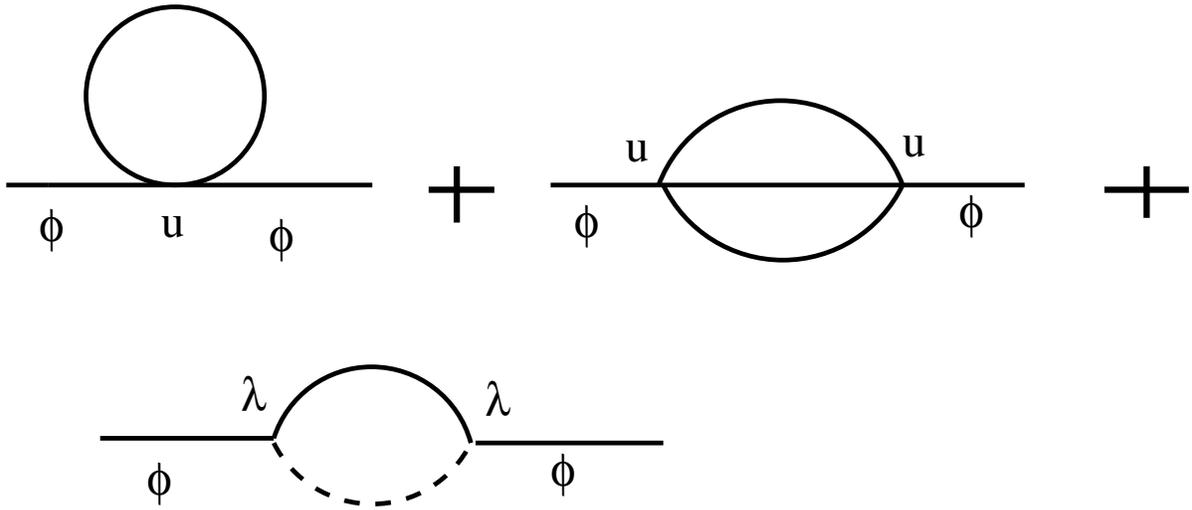}
\caption{Feynmann diagrams for $\langle\phi({\bf q})\phi(-{\bf
q})\rangle$. }
\end{figure}
\begin{figure}[htb]
\includegraphics[height=2.0cm]{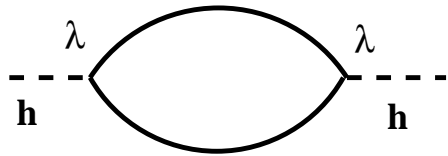}
\caption{Feynmann diagrams for $\langle h({\bf q}h(-{\bf
q}))\rangle$.}
\end{figure}
\begin{figure}[htb]
\includegraphics[height=5.0cm]{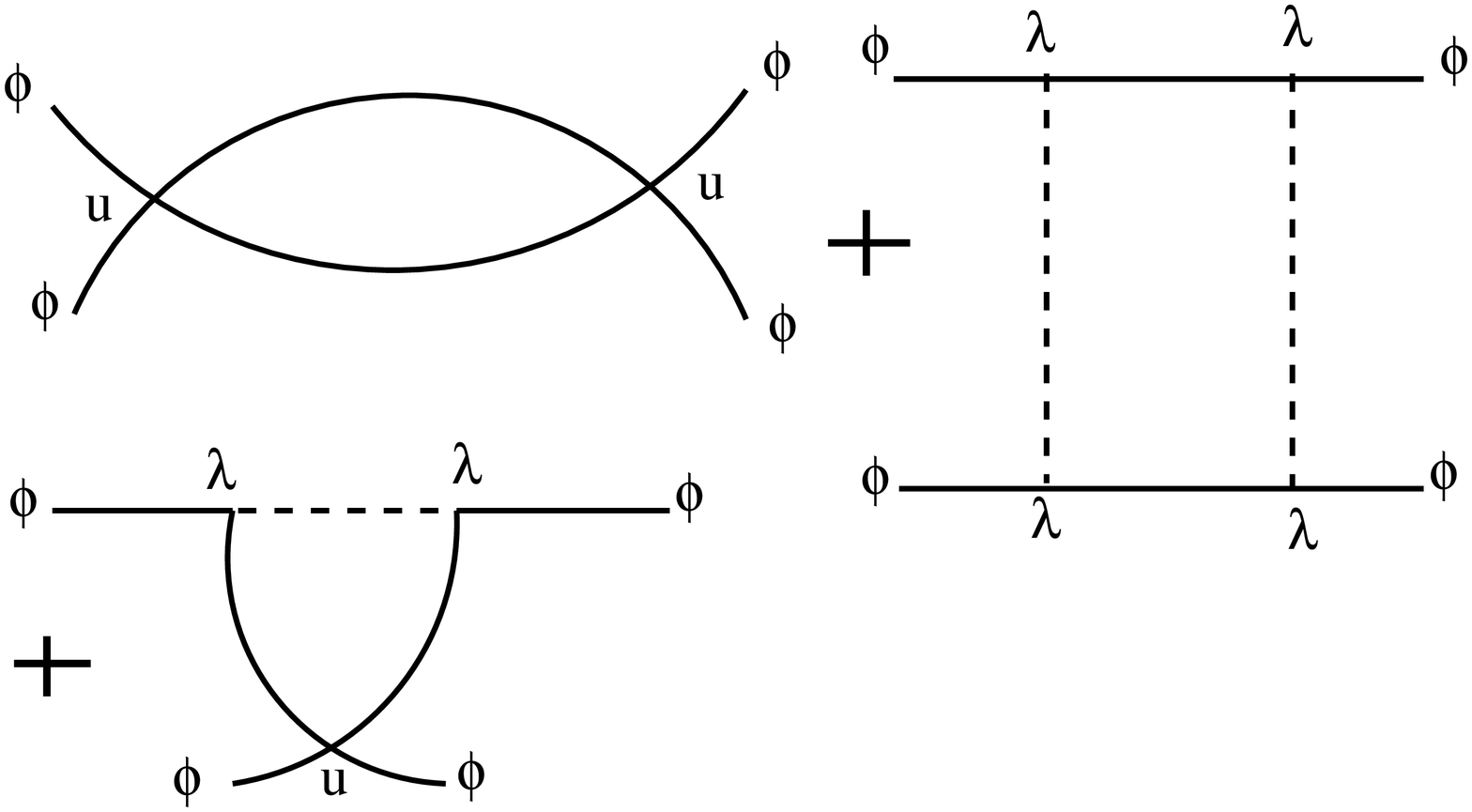}
\caption{Feynmann diagrams for $u$. }
\end{figure}
\begin{figure}[htb]
\includegraphics[height=5.0cm]{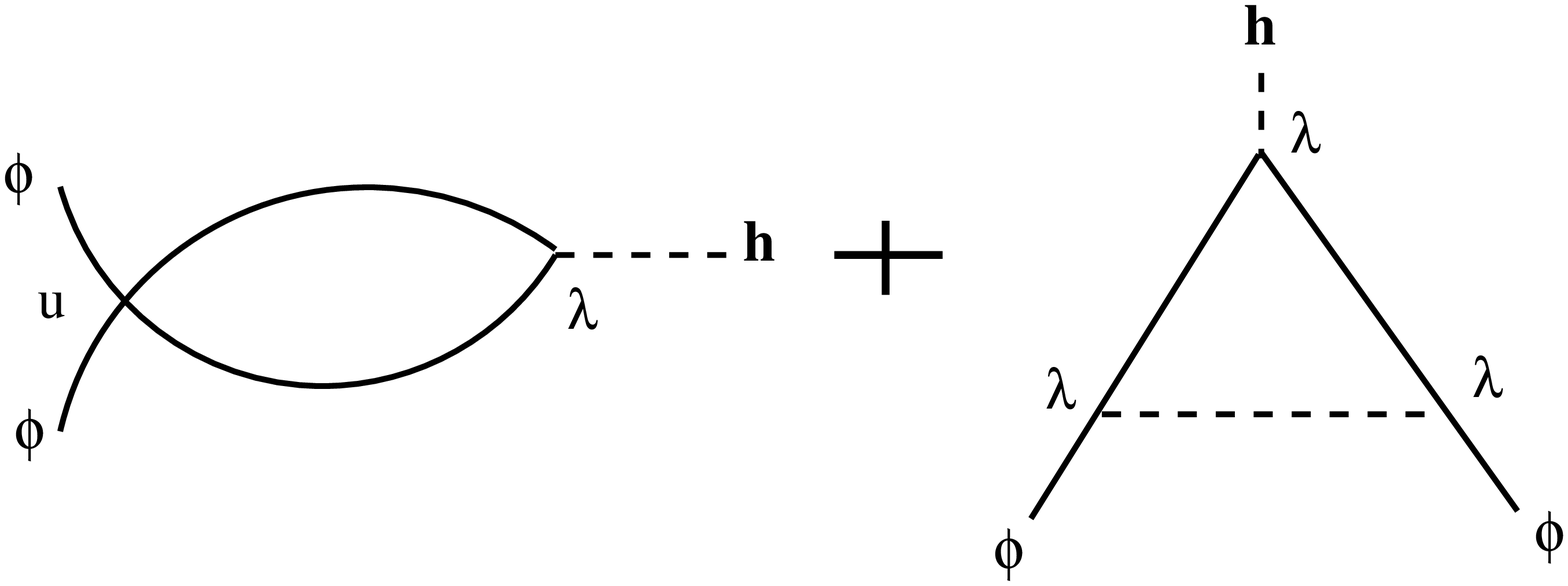}
\caption{Feynmann diagrams for $\lambda$.}
\end{figure}
\end{widetext}


\begin{thebibliography}{99}
\bibitem{hetero} See for recent reviews: F. A. Heberle and G. W.
Feigenson, {\em Cold Spring Harb Perspect Biol} {\bf 3}, 004630
(2011); K. Simons and J. L. Sampaio, {\em ibid.}, p. 4697; E. L.
Elson {\em et al}, {\em Annu. Rev. Biophys.} {\bf 39}, 207 (2010).
\bibitem{baum} T. Baumgart {\em et al}, {\em Proc. Natl. Acad. Sci. (USA)} {\bf 104}, 3165 (2007).
\bibitem{vet1} S. L. Veatch and S. L. Keller, {\em Biophys. J} {\bf
85}, 3074 (2003);
 A. R. Honerkamp-Smith, S. L. Veatch and S. L.
Keller, {\em Biochimica et Biophysica Acta} {\bf 1788}, 53 (2009).
\bibitem{crit} S. L. Veatch {\em et al}, {\em Proc. Natl. Acad. Sci. U.S. A.} {\bf 104}, 17650 (2007);
S. L Veatch, {\em ACS Chem. Biol.} {\bf 3}, 287
(2008).
\bibitem{honerkamp} A. R. Honerkamp-Smith {\em et al}, {\em Biophys.
J} {\bf 95}, 238 (2008); A. R. Honerkamp-Smith, B. B. Machta and S.
L. Keller, {\em Phys. Rev. Lett.} {\bf 108}, 265702 (2012).
\bibitem{ehrig} J. Ehrig {\em et al}, {\em New J. Ph.} {\bf 13},
045019 (2011).
\bibitem{alberts} B. Alberts, D. Bray, J. Lewis, M. Raff, K. Roberts, and J.D.
Watson, {\em Molecular Biology of the Cell}, 3rd ed. Garland, New
York (1994).
\bibitem{asym} D.W. Allender and M. Schick, {\em Biophys. J.} {\bf
91} 2928 (2006); A. J. Wagner {\em et al}, {\em Biophys. J.} {\bf 93}, 4268 (2007).
\bibitem{sym1} Since artificial lipid bilayers are generally
symmetric and some asymmetry is believed to be lost during
preparation of GPMVs, experimental results on them generally
correspond to symmetric membranes; see, e.g., Ref.~\cite{baum}
above. It is still debatable at present whether plasma
membranes of living cells undergo any phase transition.
\bibitem{linear} See for  asymmetric membrane models with linear curvature-composition
relation (i.e., invariance under SYMIII) W. Helfrich, {\em Z.
Naturforsch.} {\bf 28}, 693 (1973); S Leibler, {\em J. Physique}
{\bf 47}, 507 (1986); T. Taniguchi, {\em Phys. Rev. Lett.} {\bf 76},
4444 (1996); Y. Jiang {\em et al}, {\em Phys. Rev. E} {\bf 61}, R57
(2000); E. J. Wallace et al, Biophys J, 88, 4072 (2005); J. G\'omez,
F. Sagu\'es and R. Reigada, {\em Phys. Rev. E} {\bf 80}, 011920
(2009); T. Baumgart et al, Annu. Rev. Phys. Chem, 62, 483 (2011); C.
Zhu et al, Biophys. J., 102, 1837 (2012). In the limit of equal
compositions (which generally does not happen), the single field
model used here may be retrieved from a bilayer model with two
fields when the monolayers are {\em locked}, i.e., in the limit of
strong inter-monolayer interactions.
This is experimentally realizable~\cite{asym1}.
\bibitem{baum-nonlin} C. Zhu, S. L. Das and T. Baumgart, {\em
Biophys. J.} {\bf 102}, 1837 (2012).
\bibitem{enzyme} R. A. Moss and S. Bhattacharya, {\em J. Am. Chem. Soc.} {\bf 117}, 8688
(1995).
\bibitem{bi} D. Allender and M. Schick, {\em Biophys. J.} {\bf 91}, 2828
(2006); G. G. Putzel and M. Schick, {\em Biophys. J.} {\bf 94}, 869
(2007); A. J. Wagner, S. Loew and S May {\em Biophys. J.} {\bf 93},
4268 (2007); Y. Hirose, S. Komura and D. Andelman, {\em
ChemPhysChem} {\bf 10}, 2839 (2009).
\bibitem{surf}  The surface tension $\sigma$ of an interface saturated by surfactant
molecules is expected to vanish (see, e.g., S. A. Safran ,
Statistical Thermodynamics of Surfaces, Interfaces and Membranes
(Addison- Wesley, Reading, Mass. 1994), and thus $\sigma$ for a
fluid membrane fluctuating freely in a solvent is usually assumed to
be zero. It may be noted that biomembranes may experience
non-zero $\sigma\sim 10^{-8}N/m$, which however, becomes irrelevant
on length scales $< \sqrt{\kappa/\sigma}$.
\bibitem{weinberg-book} {\em Statistical Mechanics of Membranes and Surfaces}, edited by D. Nelson, T. Piran, and S. Weinberg World Scientific,
Singapore (1989).
\bibitem{leib1} See, e.g., S. Leibler and D. Andelman, {\em J.
Physique} {\bf 48}, 2013 (1987), for similar approaches.
\bibitem{helfrich} W. Helfrich, {\em J. Phys. (France)} {\bf 46}, 1263 (1985); H.J. Deuling and W. Helfrich, ibid. {\bf 37}, 1335 (1976).
\bibitem{wall} M. Haataja, {\em Phys. Rev. E} {\bf 80}, 020902 (R)
(2009).
\bibitem{udo-rev} See, e.g., Ref.~\cite{weinberg-book} and U.
Seifert, {\em Adv. Phys.} {\bf 46}, 13 (1997) for detailed
discussions on fluid membranes.
\bibitem{foot7} Dropping the Ising symmetry (SYMI) and not
insisting upon any other symmetry (e.g., SYMIII) allows us to have a
$\phi^3$-term even for a symmetric membrane. Since present available
experimental results point towards 2D Ising like behavior for
symmetric membranes, these considerations as above have led us to
use the Ising symmetry to build our model that uses a nonlinear
curvature-composition coupling. Note that nonlinear curvature
composition coupling of the form $\phi (\nabla^2 h)^2$ has been
considered for symmetric membranes, see, e.g., Ref.~\cite{ayton}.
This violates the Ising symmetry. In contrast, we have retained the
Ising symmetry, but broken the inversion symmetry of the membrane.
\bibitem{geom} These geometric nonlinearities originate from (i) expansion of the area element with respect to the
Monge gauge base plane for small height fluctuations and (ii)
nonlinear corrections to the form of the mean curvature in the Monge
gauge.
\bibitem{geom3} There may be an additional symmetry-allowed
inversion symmetry breaking geometric nonlinearity of the form
$\nabla^2 h ({\bf\boldsymbol \nabla}h)^2$, which may arise in an
asymmetric membrane from a linear curvature term in $\mathcal
F$: $\int dS \nabla^2 h=\int dxdy\sqrt{1+(\nabla h)^2}\nabla^2
h\approx \int dxdy (\nabla h)^2\nabla^2h$, since $\int dxdy \nabla^2
h$ does not contribute to the free energy, it being a total
derivative. While this term may be relevant (same as the $\lambda$-
or $u$-terms) we did not consider this term here for simplicity.
\bibitem{seifert1} U. Seifert and S. A. Langer, {\em Europhys.
Lett.} {\bf 23}, 71 (1993); U. Seifert and S. A. Langer, {\em
Biophys. Chem.} {\bf 49}, 13 (1994)
\bibitem{hel} E. H. Graf {\em et al}, {\em Phys. Rev. Lett.} {\bf 19}, 417 (1967).
\bibitem{chaikin} P. M. Chaikin and T. C. Lubensky, {\em Principles of condensed matter physics}, Foundation Books,
New Delhi (1998).
\bibitem{int} Formally, integrating over $\phi$ gets
problemtic near $T_c$. This can be circumvent in a RG framework.
\bibitem{desserno} I. R. Cooke, K. Kremer and M. Deserno, {\em Phys.
Rev. E} {\bf 72}, 011506 (2005).
\bibitem{geom1} Even if the geometric nonlinearity $\int d^dx \kappa (\nabla^2
h)^2 ({\boldsymbol\nabla} h)^2$ is considered, {\em renormalized} or
effetive scale-dependent $\kappa$ decreases as $\log
L$~\cite{weinberg-book}. Thus $\langle n^2 ({\bf x})\rangle$
continues to scale as $\log L$. The other possible geometric
nonlinearities are all subleading to the $\lambda$- and
$u$-nonlinearities considered here. Therefore, although for a
symmetric membrane ($\lambda=0$) all the geometric nonlearities are
present (having invariance under SYMII), their contributions to
$\Delta_s$ will not generate a $L^2$ dependence found in $\Delta$
for an asymmetric membrane in our model.
\bibitem{whyrg} Since in our model the $\phi$-field, being Ising like, has a lower critical
dimension $d_L=1$ and thus displays the standard order-disorder
transition at 2D, with a diverging correlation length near $T_c$,
formal and rigorous enumeration of the scaling exponents requires
the machinery of RG calculations.
\bibitem{amit} D. J. Amit, {\em Field Theory, the Renormalisation Group,
and Critical Phenomena} (Academic Press, NY, 1978).
\bibitem{zinn} G.  Hooft and M. Veltman, {\em Nuc. Phys. B}, {\bf 44}, 189
(1972); C. De Dominicis, E. Br\'ezin and J. Zinn-Justin, {\em Phys.
Rev. B}, {\bf 12}, 4945 (1975); E. Br\'ezin, J. C. L. Guillon, and
J. Zinn-Justin, {\em Phase Transitions and Critical Phenomena}, vol.
6 (Academic Press, NY, 1976); J. Zinn-Justin, {\em Quantum Field
Theory and Critical Phenomena}  (Oxford Science Publication, Oxford
2010).
\bibitem{loop}  Fluctuation corrections to coupling constants $u$ and $\lambda$ are evaluated
up to the one-loop order. If higher order loop corrections are
included, then corrections to the exponents higher order in
$\epsilon$ should arise. However, we do not forsee appearance of any
new divergence that cannot be absorbed by renormalizing the existing
model parameters and fields. See Appendix for the relevant Feynmann diagrams.
\bibitem{asym1}  M. D. Collins and S. L. Keller, {\em Proc.
Natl. Acad. Sci. USA} {\bf 105}, 124 (2008).
\bibitem{ayton} G. S. Ayton {\em et al}, {\em Biophys. J.} {\bf 88},
3855 (2005).
\bibitem{lang} J. Liu and J. C. Conboy, {\em Langmuir} {\bf 21}, 9091 (2005); J. Yuan, C. Hao, M. Chen, P. Berini and S. Zou, {\em
Langmuir} {\bf 29}, 221 (2013).
\bibitem{flicker} T. Betz and C. Sykes, {\em Soft Matter} {\bf 8},
5317 (2012).
\bibitem{exp} See, e.g., J. Song and R. E. Waugh, {\em Biophys. J} {\bf 64}, 1967
(1993); L. R. Arriaga {\em et al}, {Biophys. J.} {\bf 96}, 3629
(2009);  C. L. Armstrong {\em et al}, {Eur. Biophys. J}, (2012) for
studies on how cholesterol may be used to tune membrane bending
rigidity. See also A. Yaghmur {\em et al}, {\em Plos one}, e479
(2007) for studies on tuning curvature by designer lipid-like
peptide surfactants. Additionally, BAR proteins are known to
rigidify membranes, see, e.g., J. Gilden and M. F. Krummel, {\em
Cytoskeleton (Hoboken)} {\bf 67}, 477 (2010).
\bibitem{kumar} P. B. S. Kumar {\em et al}, {\em Phys. Rev. Lett.}  {\bf 86}, 3911
(2001).

\end{thebibliography}
\end{document}